\documentclass[iop,useAMS,usenatbib]{emulateapj}

\usepackage{graphicx}
\usepackage[percent]{overpic}
\usepackage{multirow}
\usepackage{color}
\usepackage{rotating}
\usepackage{amsmath}
\usepackage{mathtools}
\usepackage{physics}

\newcommand{\msun}{${\rm M_{\sun}}$}

\def\ltsima{$\; \buildrel < \over \sim \;$}
\def\simlt{\lower.5ex\hbox{\ltsima}}
\def\gtsima{$\; \buildrel > \over \sim \;$}
\def\simgt{\lower.5ex\hbox{\gtsima}}
\def\kms{{\rm\,km\,s^{-1}}}
\def\mas{{\rm\,mas}}

\def\kpc{{\rm\,kpc}}

\def\msun{{\rm\,M_\odot}}

\def\pc{{\rm\,pc}}

\def\deg{^\circ}
\def\s{\ifmmode \widetilde \else \~\fi}
\def\={\overline}

\def\spose#1{\hbox to 0pt{#1\hss}}

\def\lta{\mathrel{\spose{\lower 3pt\hbox{$\mathchar"218$}}
     \raise 2.0pt\hbox{$\mathchar"13C$}}}
\def\gta{\mathrel{\spose{\lower 3pt\hbox{$\mathchar"218$}}
     \raise 2.0pt\hbox{$\mathchar"13E$}}}
\def\Dt{\spose{\raise 1.5ex\hbox{\hskip3pt$\mathchar"201$}}}    % upper case
\def\dt{\spose{\raise 1.0ex\hbox{\hskip2pt$\mathchar"201$}}}    % lower case

\def\dotsfill{\leaders\hbox to 1em{\hss.\hss}\hfill}

\def\Gyr{{\rm\,Gyr}}

\def\ltsima{$\; \buildrel < \over \sim \;$}
\def\gtsima{$\; \buildrel > \over \sim \;$}
\def\lsim{\lower.5ex\hbox{\ltsima}}
\def\gsim{\lower.5ex\hbox{\gtsima}}
\def\lapp{\ifmmode\stackrel{<}{_{\sim}}\else$\stackrel{<}{_{\sim}}$\fi}
\def\gapp{\ifmmode\stackrel{>}{_{\sim}}\else$\stackrel{<}{_{\sim}}$\fi}

\shorttitle{The Phlegethon stellar stream}
\shortauthors{Ibata, Malhan, Martin \& Starkenburg}

\begin{document}

\title{Phlegethon, a nearby $75\deg$-long retrograde stellar stream}

\author{Rodrigo A. Ibata\altaffilmark{1}}
\author{Khyati Malhan\altaffilmark{1}}
\author{Nicolas F. Martin\altaffilmark{1,2}}
\author{Else Starkenburg\altaffilmark{3}}

\altaffiltext{1}{Observatoire Astronomique, Universit\'e de Strasbourg, CNRS, 11, rue de l'Universit\'e, F-67000 Strasbourg, France; rodrigo.ibata@astro.unistra.fr}

\altaffiltext{2}{Max-Planck-Institut f\"ur Astronomie, K\"onigstuhl 17, D-69117 Heidelberg, Germany}

\altaffiltext{3}{Leibniz Institute for Astrophyics Potsdam (AIP), An der Sternwarte 16, D-14482 Potsdam, Germany}

\begin{abstract}
We report the discovery of a $75\deg$ long stellar stream in Gaia DR2 catalog, found using the new \texttt{STREAMFINDER} algorithm. The structure is probably the remnant of a now fully disrupted globular cluster, lies $\approx 3.8\kpc$ away from the Sun in the direction of the Galactic bulge, and possesses highly retrograde motion. We find that the system orbits close to the Galactic plane at Galactocentric distances between $4.9$ and $19.8\kpc$. The discovery of this extended and extremely low surface brightness stream ($\Sigma_G\sim 34.3 \, {\rm mag \, arcsec^{-2}}$) with a mass of only $2580\pm140\msun$, demonstrates the power of the \texttt{STREAMFINDER} algorithm to detect even very nearby and ultra-faint structures. Due to its proximity and length we expect that Phlegethon will be a very useful probe of the Galactic acceleration field.
\end{abstract}


\section{Introduction}
\label{sec:Introduction}

The arrival of the second data release (DR2) of the Gaia mission has opened up the field of Galactic Archeology to exciting new endeavors that were previously completely out of reach. The excellent parallax and proper motion measurements \citep{2018arXiv180409366L} of over a billion stars now allow the the dynamics of the various constituents of our Galaxy to be studied in great detail \citep{2018arXiv180409372K,2018arXiv180409381G}, enabling progress towards the ultimate goal of understanding the formation of our Galaxy, its stellar components and the dark matter.

Among the Galactic components, stellar streams account for only a very minor fraction of the total mass budget, yet their astrophysical interest far exceeds their small contribution to our Galaxy. This importance stems partly from the fact that they represent fossil remnants of the accretion events that built up the Milky Way, giving us a means to ascertain the number and provenance of the building blocks of the Galactic halo \citep{Johnston:2008jp}. Streams also roughly delineate orbits in the Galaxy, allowing one to probe the gradient of the gravitational potential \citep{Ibata:2001be,2010ApJ...714..229L,2011MNRAS.417..198V,2015ApJ...803...80K,2016ApJ...833...31B,2018arXiv180406854B} in a way that is independent of methods that make the often-unjustified assumption of dynamical equilibrium of a tracer population. Furthermore, the low velocity dispersion and fine transverse width of low-mass streams renders them excellent probes of the small-scale substructure of the dark matter halo \citep{2002MNRAS.332..915I,2002ApJ...570..656J,2012ApJ...748...20C,2016MNRAS.463..102E}.

In a previous contribution, we introduced a new algorithm (the \texttt{STREAMFINDER}; \citealt[][hereafter paper~I]{2018MNRAS.477.4063M}) built specifically to search efficiently through astrometric and photometric databases for stream-like structures. The first results of this algorithm applied to Gaia DR2 were presented in \citet[][hereafter paper~II]{2018arXiv180411339M}, but were limited to distances $>5\kpc$, a choice that we made in order to reduce the necessary calculation time. Here we describe a discovery based upon re-running the algorithm searching for stellar streams at distances between 0.5 and $5\kpc$. 

The layout of this article is as follows. Section~\ref{sec:Data} explains the selection criteria that were applied to the data and briefly summarizes the detection algorithm and adopted parameters. The results of the analysis are reported in Section~\ref{sec:Results}. We present a refined fit to the stream in Section~\ref{sec:Orbital_Fits}, and our discussion and conclusions in Section~\ref{sec:Discussion}.

\section{Data and Stream Analysis}
\label{sec:Data}

All the data used in the present analysis were drawn from the Gaia DR2 catalog \citep{2016A&A...595A...1G,2018arXiv180409365G}. As described in paper~II, we extinction-corrected the survey using the \citet{Schlegel:1998fw} dust maps, and kept only those stars with $G_0<19.5$; this limiting magnitude ensures a homogeneous depth over the sky, given the additional selection criterion of choosing to analyze the sky at $|b|>20\deg$. 

A full exposition of the updated \texttt{STREAMFINDER} algorithm and contamination model will be provided in Ibata et al. (2018, in prep), which includes a demonstration that real structures are easily distinguishable from false positives. Briefly, the algorithm works by examining every star in the survey in turn, sampling the possible orbits consistent with the observed photometry and kinematics, and finding the most likely stream fraction given a contamination model and a stream model. The adopted contamination model is identical to that built for paper~II. It is an empirical model in sky position, color-magnitude and proper motion space (i.e. six-dimensions) constructed by spatially randomizing the Gaia counts with a $2\deg$ Gaussian. While the correlations between sky position and color-magnitude are recorded as binned arrays, the color-magnitude and proper motion information is condensed by the use of a 100-component Gaussian mixture model in each spatial bin.

\begin{figure}
\begin{center}
\includegraphics[angle=0, viewport= 45 30 500 520, clip, width=\hsize]{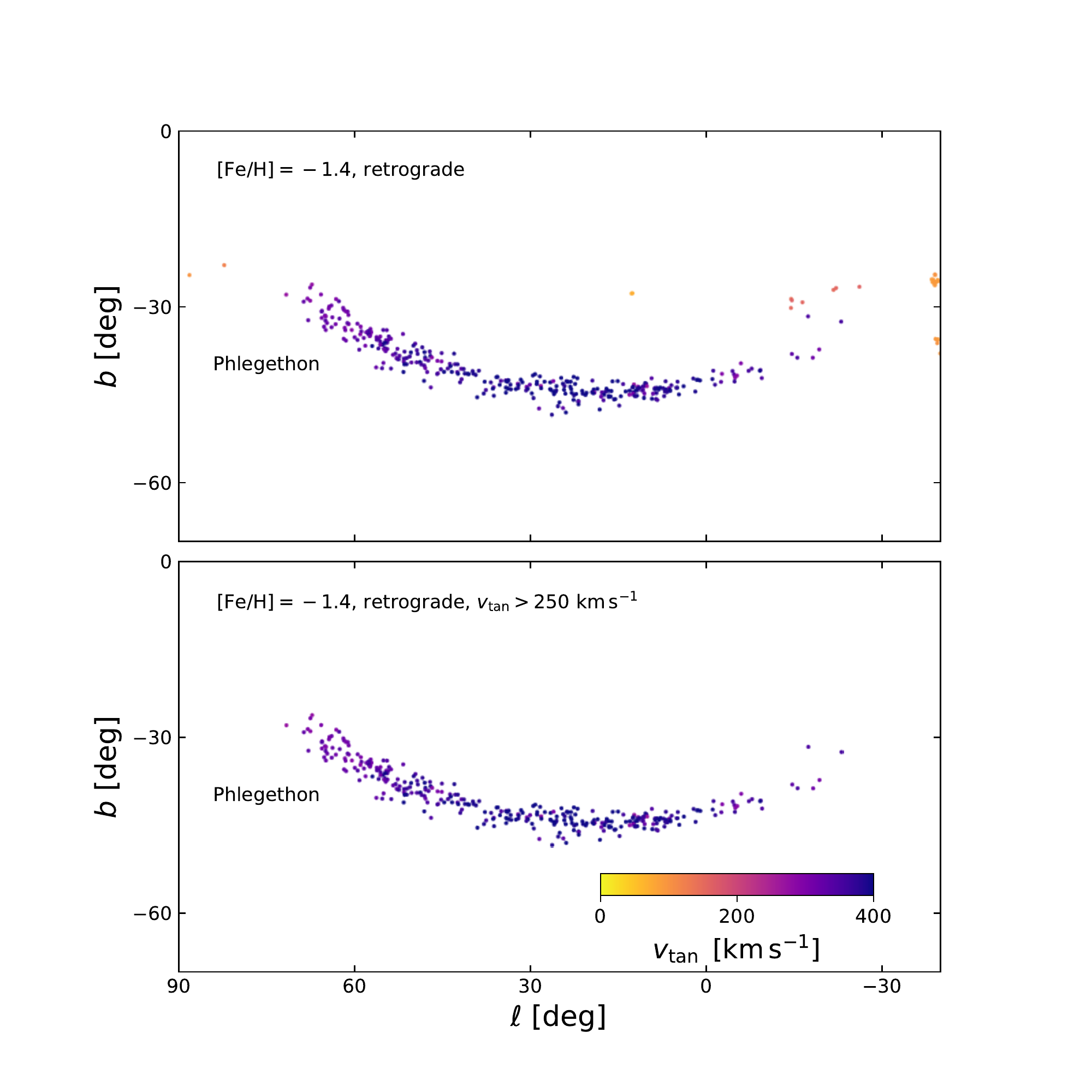}
\end{center}
\caption{Output of the {\tt STREAMFINDER} software in the distance range $[0.5, 5.0]\kpc$ over the region of sky $-40\deg< \ell < 90\deg$, $-70\deg<b<-20\deg$, assuming a stellar population of age $10\Gyr$ and ${\rm [Fe/H]=-1.4}$. Only those stars are plotted that have retrograde orbital solutions, and for which the stream detection significance is greater than $8\sigma$. The upper panel shows the full range of tangential motions, while the bottom panel retains only those stars with $v_{\rm tan}>250\kms$. A clear stream, labelled ``Phlegethon'', is detected as a continuous linear structure in the panels. The color of the points marks the tangential velocity of the stars, calculated using the observed Gaia proper motion in conjunction with the distance solutions that the algorithm derives for every processed star.}
\label{fig:FeH_m1.4}
\end{figure}

The adopted stream model is very simple: we integrate along the sampled orbits for a half-length of $10\deg$ (i.e. total length $20\deg$), and over this length the stream counts have uniform probability. Perpendicular to the stream model, the properties of all observables are taken to be Gaussian, so the model has a Gaussian physical width (selected here to be $100\pc$), a Gaussian dispersion in both proper motion directions equivalent to $3\kms$, and a dispersion in distance modulus of $0.05\,{\rm mag}$. These width and velocity dispersion parameters are similar to the properties of known globular cluster streams such as the Palomar~5 stream \citep{Dehnen:2004ez,2016ApJ...819....1I}, while the distance modulus dispersion was adopted to allow for a small mismatch between the color-magnitude behavior of the adopted stellar populations model and that of the real stellar population.

The {\tt STREAMFINDER} requires a model of the Galactic potential in order to calculate orbits; for this we use the realistic Milky Way mass model of \citet{Dehnen:1998tk}, their model `1'. The algorithm returns the best-fit orbit out of the sampled set for a given data point, the number of stars in the corresponding putative stream and the likelihood value relative to the model where the stream fraction is zero. All stars shown below have likelihoods equivalent to a stream detection exceeding the $8\sigma$ level.

To convert the kinematics of the integrated orbits into the space of observables, we assume that the Sun lies $8.20\kpc$ from the Galactic center and $17\pc$ above the Galactic mid-plane \citep{2017MNRAS.465..472K}, that the local circular velocity is $V_{\rm circ} = 240\kms$ and that the peculiar velocity of the Sun is $(u_{\odot}, v_{\odot}, w_{\odot}) = (9.0, 15.2, 7.0) \kms$ \citep{2014ApJ...783..130R, 2010MNRAS.403.1829S}.

The main aim in developing the \texttt{STREAMFINDER} was to try to find distant halo streams for which the Gaia parallax measurements are poor. To circumvent this deficiency, we use stellar populations models to convert the measured photometry into trial line of sight distance values. To this end we adopted the PARSEC isochrone models 
\citep{2012MNRAS.427..127B}, covering a range in age and metallicity. Here, however, we report results using a single model with age $10\Gyr$ and metallicity of ${\rm [Fe/H]=-1.4}$.

The stream solutions found by the \texttt{STREAMFINDER} automatically include the best-fit orbit over the trial stream length, from which one can naturally obtain an estimate of the sense of rotation of the survey stars with respect to the Galaxy. This is possible because the algorithm finds a value for the missing line of sight velocity information of a star in the Gaia DR2 catalog by requiring continuity to nearby stream candidates. The radial velocity is sampled between $\pm$ the escape velocity, and the value corresponding the the maximum-likelihood stream solution is retained. We showed in paper~II that this works for the case of well-measured streams such as GD-1 \citep{2006ApJ...643L..17G}. We use this orbital information to classify stars as either prograde or retrograde.

\begin{figure}
\begin{center}
\includegraphics[angle=0, viewport= 10 10 395 390, clip, width=\hsize]{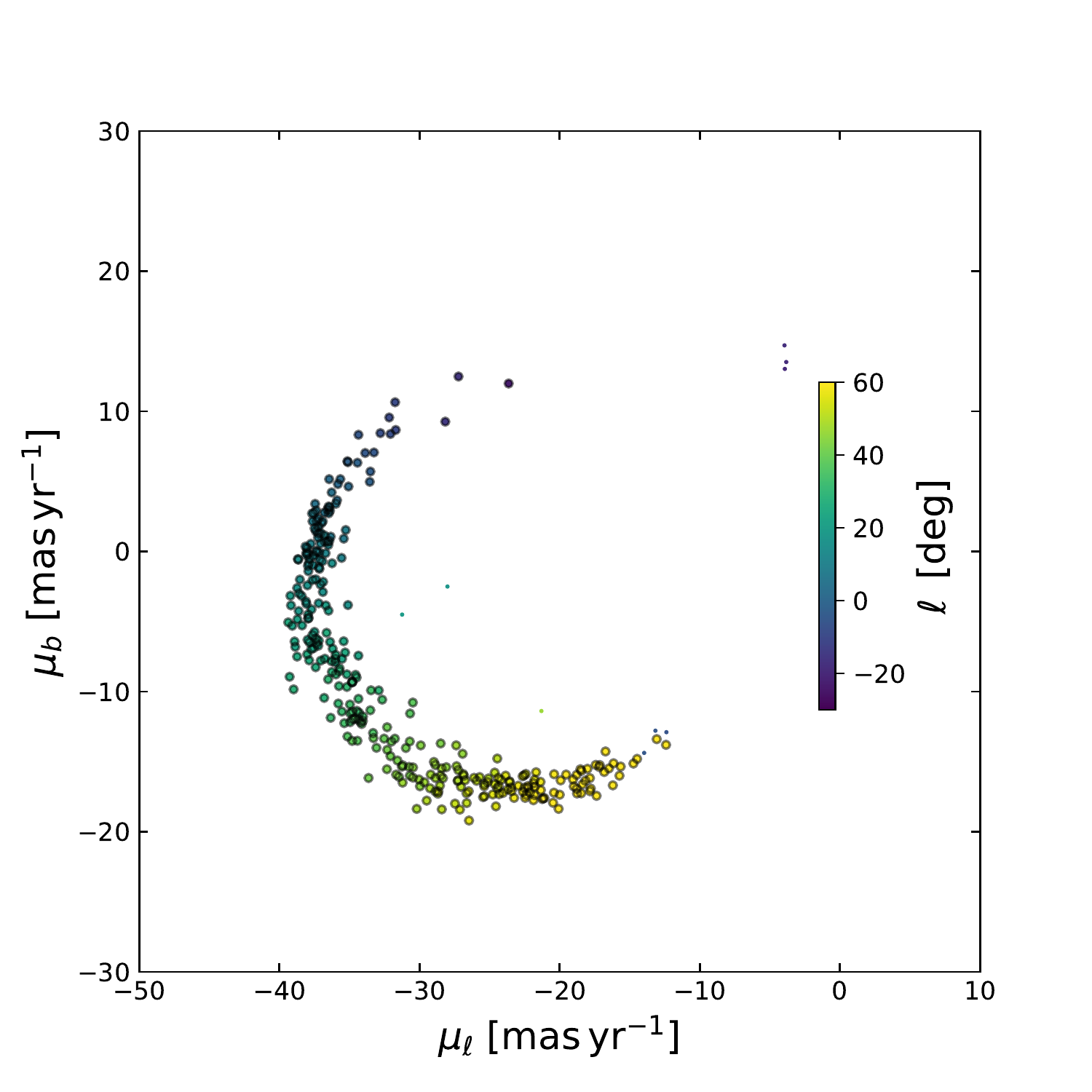}
\end{center}
\caption{Proper motion distribution of the sample of stars in the lower panel of Figure~\ref{fig:FeH_m1.4}. The color of the points marks Galactic longitude, which can be seen to vary continuously along the arc-shaped feature in this distribution. This distinct structure was selected to lie within an  irregular hand-drawn polygon in this proper motion plane; the resulting sample is shown with the larger (black circled) dots.}
\label{fig:PM_distribution}
\end{figure}

\begin{figure}
\begin{center}
\includegraphics[angle=0, viewport= 15 15 505 450, clip, width=\hsize]{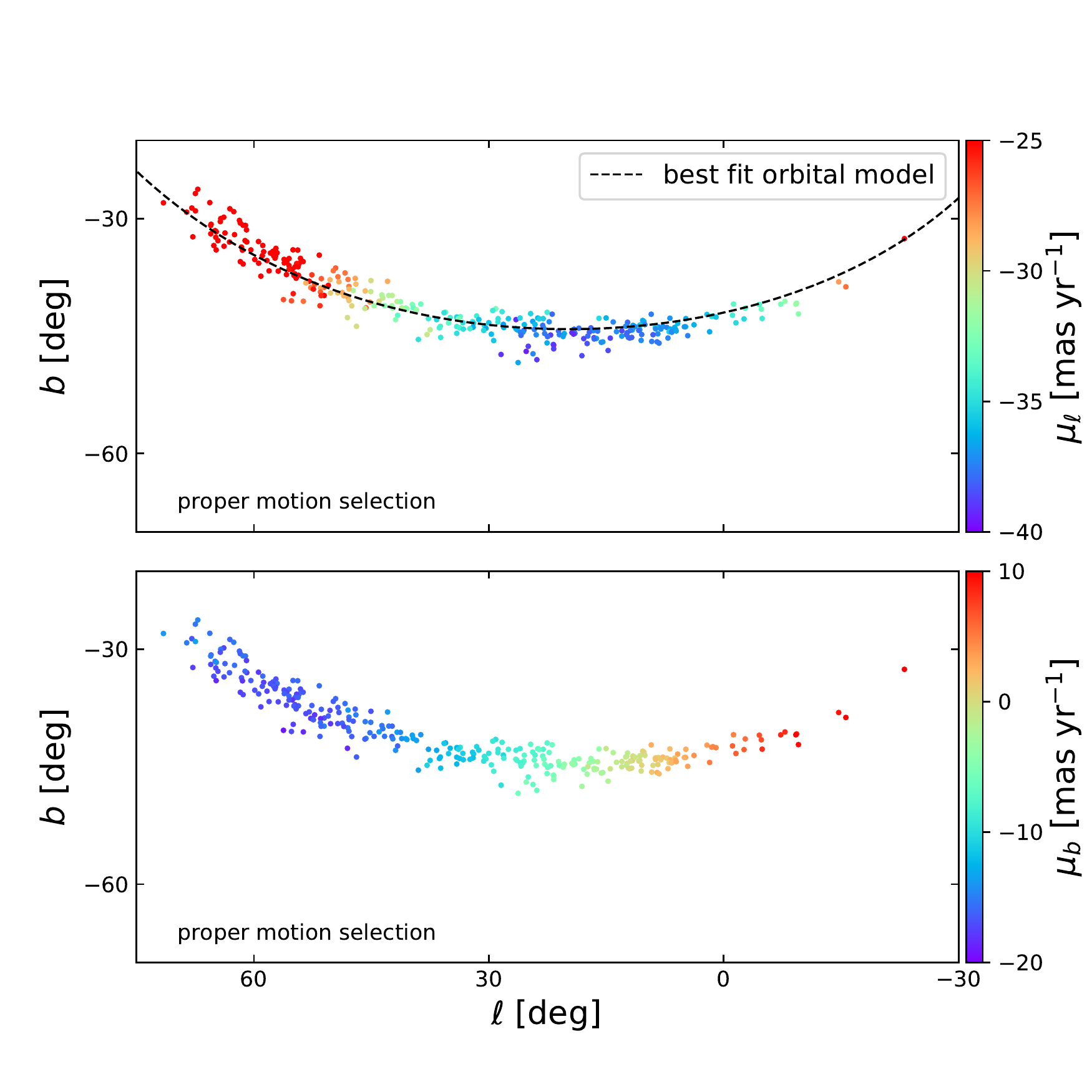}
\end{center}
\caption{Sky distribution of the selection of 321 stars drawn from Figure~\ref{fig:PM_distribution}. A very clean stream-like structure is revealed by the proper motion selection. The proper motion distribution along the stream in Galactic $\mu_\ell$ and $\mu_b$ are color-coded on the top and bottom panels, respectively. The continuity of the proper motion distribution is directly obvious from the individual proper motion measurements. The dashed line in the upper panel shows the path of the best fit orbit described in Section~\ref{sec:Orbital_Fits}.}
\label{fig:sky_plots}
\end{figure}

\section{Results}
\label{sec:Results}

In Figure~\ref{fig:FeH_m1.4} (top panel) we show the 358 stars in the Gaia DR2 catalog for which the \texttt{STREAMFINDER} solutions are retrograde in the area of sky $-40\deg< \ell < 90\deg$, $-70\deg<b<-20\deg$. The parameter choices discussed previously were used. The color of the points encodes the tangential velocity $v_{\rm tan}$ of the stars, derived from the measured proper motions and the distances estimated by the algorithm. A stream feature is clearly present in this retrograde sample, which becomes even more evident when selecting only those stars with $v_{\rm tan}> 250\kms$; this filtering yields 330 stars, shown on the bottom panel. We name this stream ``Phlegethon'', after the river of the Greek Underworld. This structure forms a $\sim 75\deg$-long arc, skirting, in projection, the southern regions of the Bulge.

\begin{figure}
\begin{center}
\includegraphics[angle=0, viewport= 10 10 395 390, clip, width=\hsize]{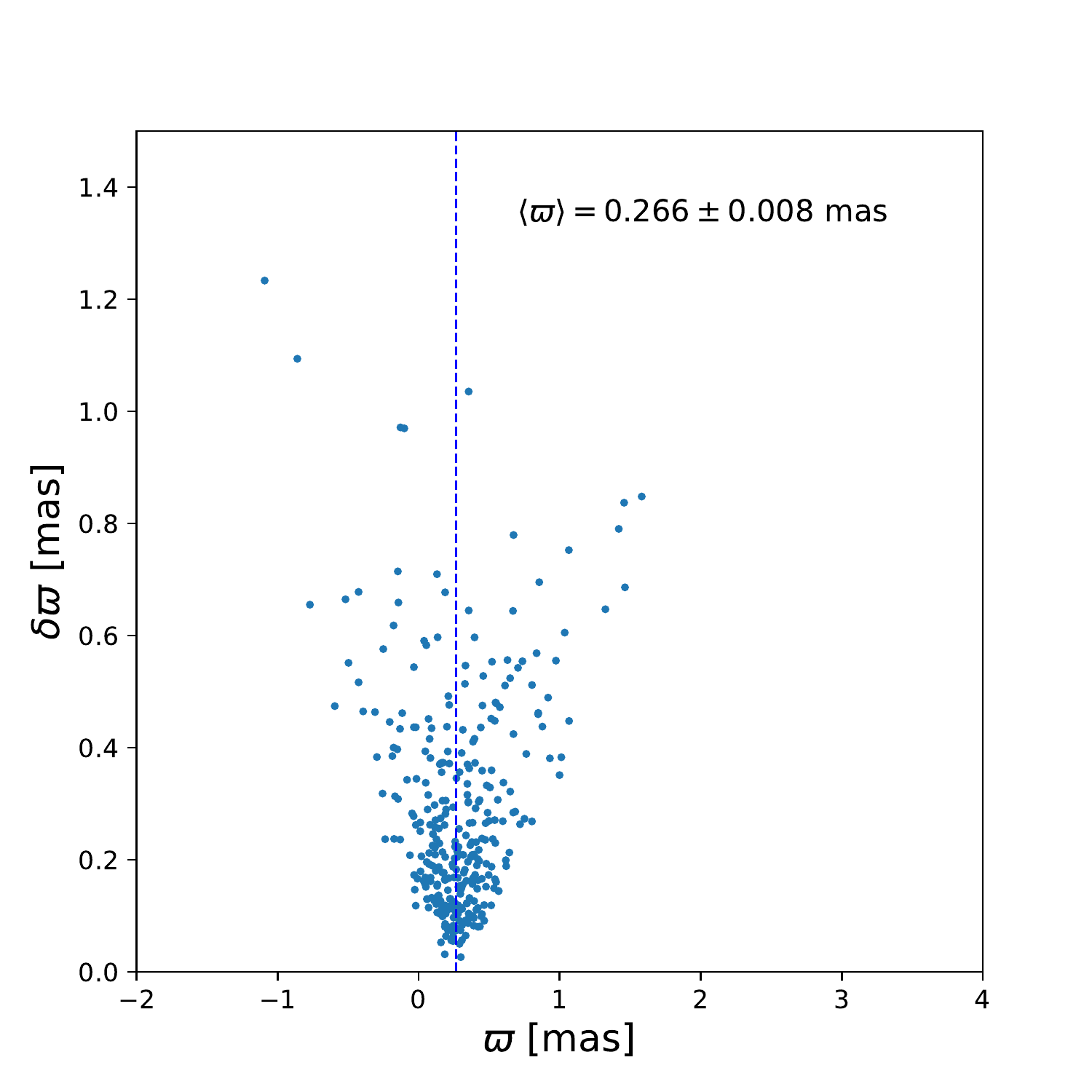}
\end{center}
\caption{Parallax distribution of the stars selected in Figure~\ref{fig:PM_distribution}. A maximum-likelihood fit to these 321 stars gives a mean parallax of $\langle \varpi \rangle=0.266\pm0.008 \mas$, implying that the structure lies at a mean distance of $\sim 3.8\kpc$.}
\label{fig:parallax}
\end{figure}

In Figure~\ref{fig:PM_distribution} we display the proper motion distribution of this sample of 330 candidate stream stars, which reveals a continuous structure with a very large gradient in proper motion. The arc-like feature is highlighted with large (circled) dots, which are colored according to Galactic longitude; the continuous variation in color demonstrates the coherence in these three parameters. These stars were selected by simply drawing an irregular polygon in the $\mu_\ell$, $\mu_b$ plane around this visually-obvious grouping. The distribution on the sky of the 321 sources selected in that arc are shown in Figure~\ref{fig:sky_plots}, which reveals a very clean stream-like feature. The proper motions in $\mu_\ell$ (top panel) are strongly negative, meaning that the stars move towards the right in the Figure. Since the distance solutions provided by the algorithm predict that the structure is located at a mean heliocentric distance of $3.57\pm0.28\kpc$, the stars must be strongly lagging the Sun. Indeed, towards $\ell=15\deg$, the proper motion in the longitude direction reaches $\mu_\ell = -38 \, {\rm mas \, yr^{-1}}$, implying a tangential velocity of $v_{\rm tan} \sim 640\kms$ with respect to the observer. The proper motions in the Galactic latitude direction $\mu_b$ (bottom panel) can be seen to be consistent with motion away from the plane moving rightwards from $\ell=70\deg$, reaching $\mu_b=0$ at $\ell \sim 15\deg$, and then the stars start moving towards the Galactic plane at $\ell<15\deg$.

\begin{figure}
\begin{center}
\includegraphics[angle=0, viewport= 10 10 395 390, clip, width=\hsize]{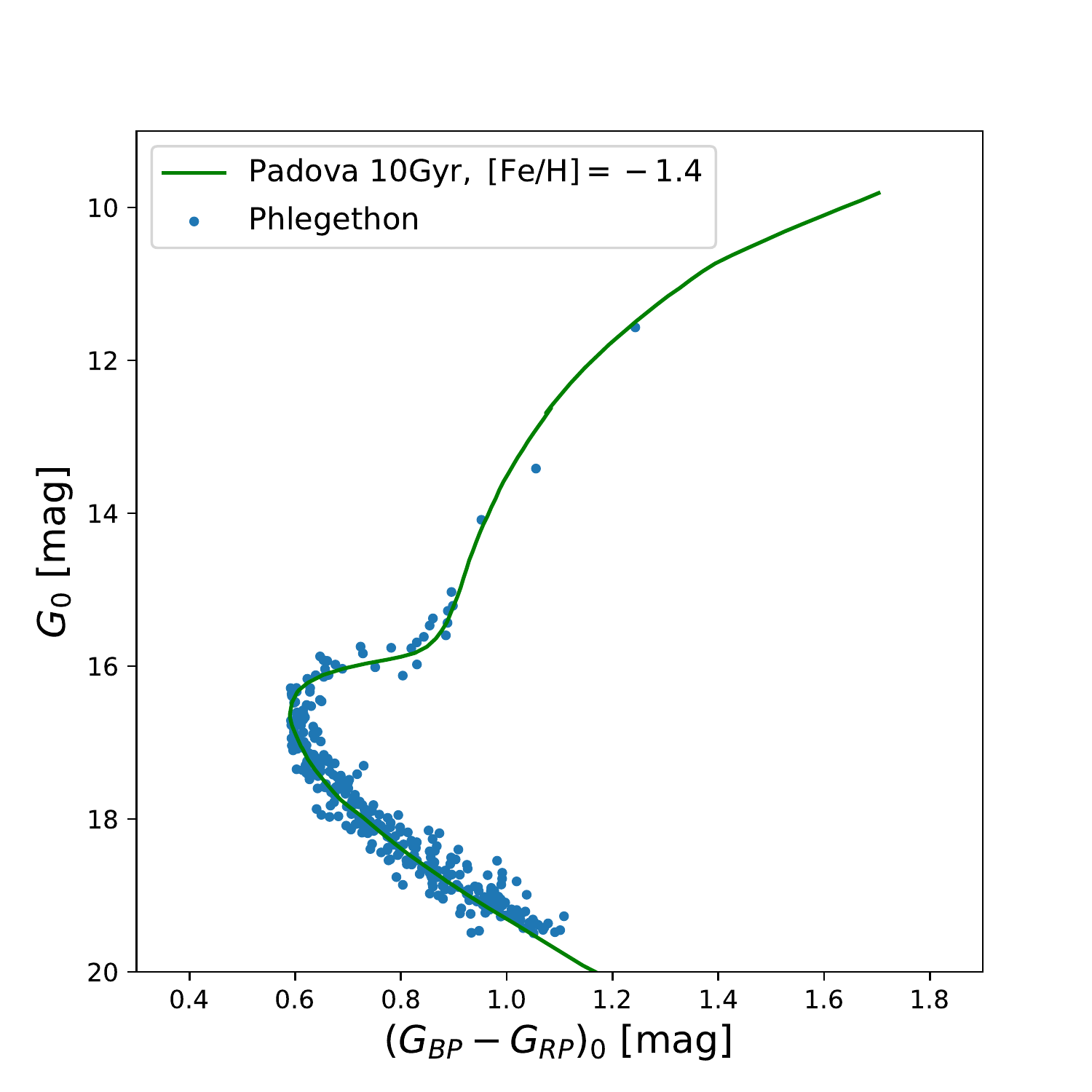}
\end{center}
\caption{Color-magnitude distribution of the sample selected in Figure~\ref{fig:PM_distribution}. The dots show the de-reddened Gaia $G$ vs. $G_{\rm BP}-G_{\rm RP}$ photometry of the stars in the stream. By construction, these should follow the chosen template stellar population model (green line), but it is nevertheless a useful check to verify that the stars are not clumped in an unphysical way on the CMD. The Padova model shown here has been shifted to account for a distance modulus of $12.88~{\rm mag}$, as measured in Figure~\ref{fig:parallax}.}
\label{fig:CMD}
\end{figure}

\begin{figure}
\begin{center}
\includegraphics[angle=0, viewport= 52 1 590 325, clip, width=\hsize]{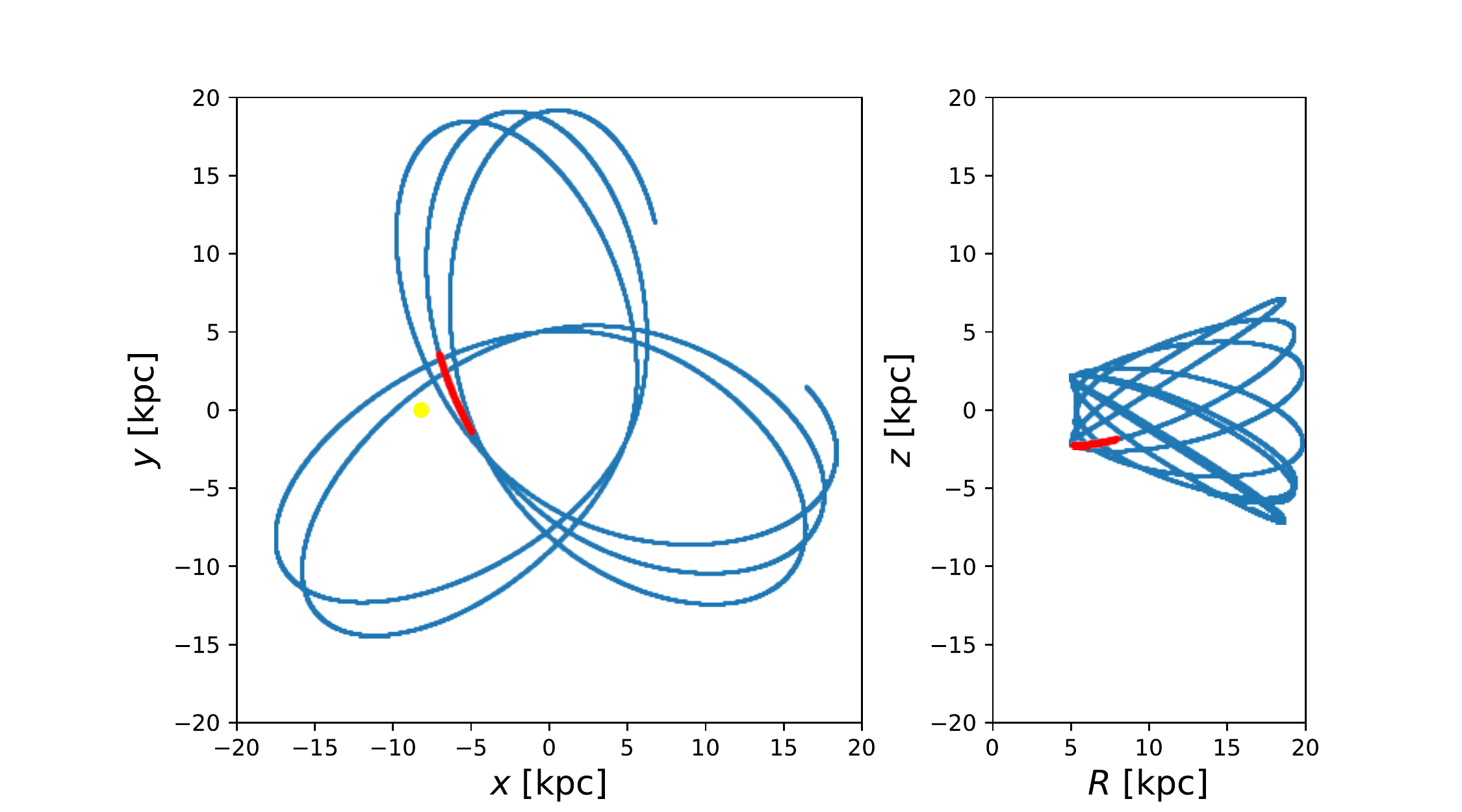}
\end{center}
\caption{The orbital path of the Phlegethon stream. The MCMC procedure described in the text was used to fit the full sample simultaneously, yielding the best-fit orbit shown here. The $x-y$ plane is shown on the left panel, and the $R-z$ on the right panel, with the blue dots showing the path $1\Gyr$ into the past and $1\Gyr$ into the future. The short red region corresponds to the area where the observed stream stars are currently located. The progenitor clearly stayed close to the plane of the Galaxy on a retrograde tube orbit. In this Cartesian system, the Galactic center is at the origin, and the Sun (marked with a large yellow dot) lies at $(x,y,z)=(-8.2,0,0.017)\kpc$.}
\label{fig:orbit}
\end{figure}

If the structure is really as close as the \texttt{STREAMFINDER} software predicts it to be, its distance should be easily resolvable with Gaia's parallax measurements. We display the parallaxes $\varpi$, along with their uncertainties $\delta\varpi$ in Figure~\ref{fig:parallax}. We implemented a simple maximum likelihood algorithm to calculate the mean parallax, assuming individual Gaussian parallax uncertainties on each star. The mean parallax was found to be $\langle \varpi \rangle=0.266\pm0.008 \mas$, i.e. $\sim 3.8\kpc$ in distance, in excellent agreement with the mean distance of the \texttt{STREAMFINDER} solutions. 

The color-magnitude distribution of the 321 stars in the sample is displayed in Figure~\ref{fig:CMD}. These can be seen to conform well to the input template stellar population model, with most stars lying on the main sequence, and just a handful on the lower red giant branch. Given that the stream is approximately $75\deg$ long and $4\deg$ wide, we deduce a system surface brightness of $\Sigma_G=34.3 \, {\rm mag \, arcsec^{-2}}$. Adopting the template stellar population model, and accounting for the fainter (unobserved) stars, implies a mass of $2580\pm140\msun$ over the observed region, where we have assumed that the Gaia DR2 is complete to $G_0=19.5$.

By good fortune (and after the initial submission of this contribution), we realised that two stars out of the 321 candidate stream stars selected in Figure~\ref{fig:PM_distribution} have spectroscopic observations in the SDSS/Segue  (DR10) survey \citep{2009AJ....137.4377Y}. These two members (with SDSS ``specobjid'' identifiers: 1712619976675846144 and 2210282052404747264) are shown on the bottom panel of Figure~\ref{fig:orbit-data}. These objects have heliocentric radial velocities of $-318.4 \pm 10.4\kms$ and $-301.2 \pm 4.6\kms$, respectively. They both lie very close to one another ($\ell=54.4\deg$ and $\ell=54.2\deg$, respectively). While the metallicity of the first star is unconstrained, the second star  has a measured spectroscopic metallicity of ${\rm [Fe/H]}=-1.56\pm0.04$, similar in value to the stellar populations template used here.

\section{Refined Orbital Fits}
\label{sec:Orbital_Fits}

Although \texttt{STREAMFINDER}  fits the orbits of the putative streams, it is designed for stream detection rather than careful fitting. We therefore updated the Lagrange-point stripping method described in \citet{2011MNRAS.417..198V}, but now using the measured proper motions and parallax information as constraints. The measured radial velocities of the two SDSS stars were also used in the fitting procedure. We dealt with the missing radial velocity information for the remaining 319 stars by assigning them all a velocity of zero, but with a Gaussian uncertainty of $10^4\kms$. We also updated the fitting procedure to use a custom-made MCMC driver package, discussed previously in \citet{2013MNRAS.428.3648I}, that uses the affine-invariant ensemble sampling method of \citet{Goodman:2010we}. Since we do not know the location of the progenitor remnant, we ignored the self-gravity of the stream in the modeling, i.e. we fitted orbits over the full length of the detected structure.

\begin{figure}
\begin{center}
\includegraphics[angle=0, viewport= 5 5 570 860, clip, width=\hsize]{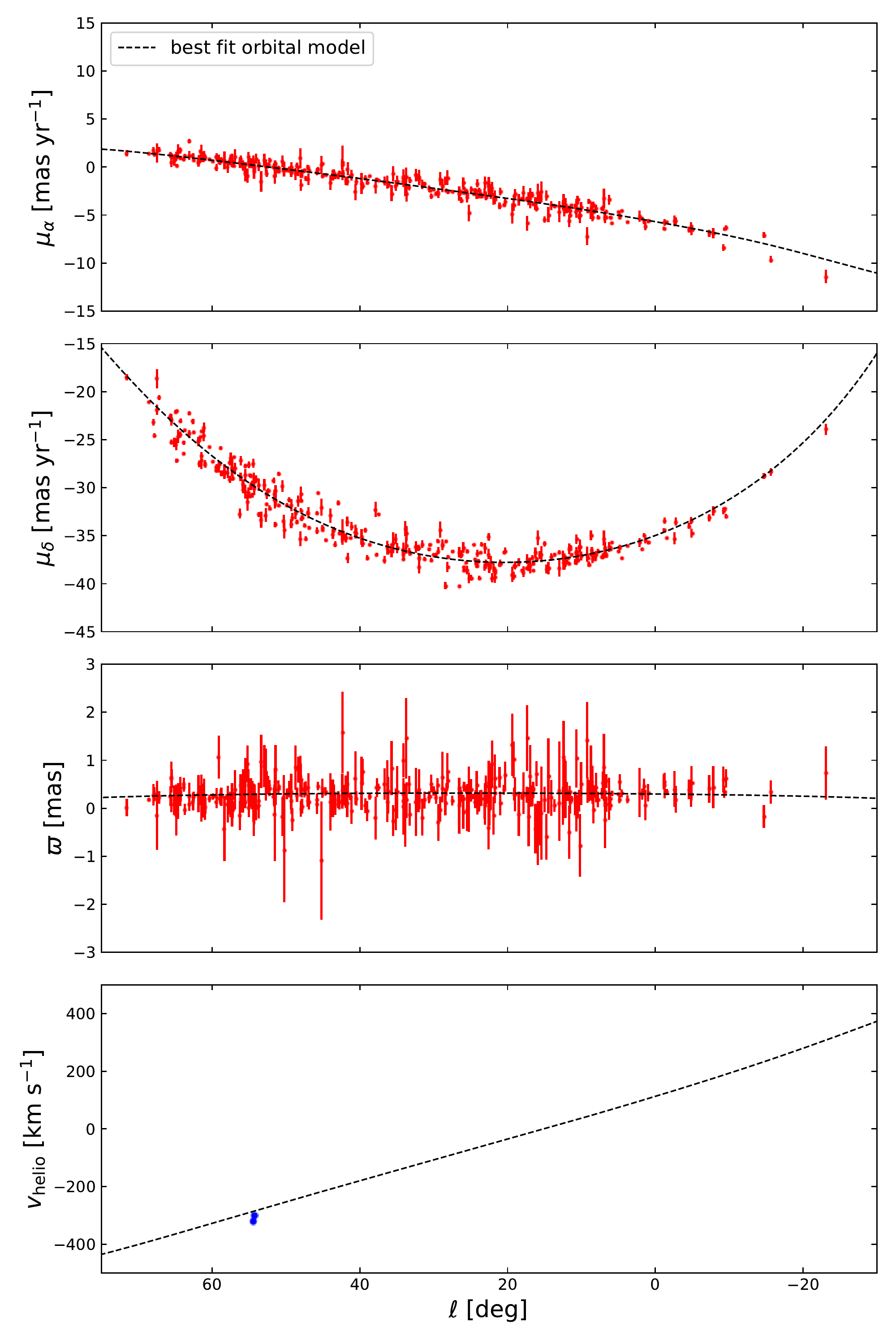}
\end{center}
\caption{Comparison of the orbital solution (black dashed lines) to Gaia DR2 data (plotted in red along with their observational uncertainties). The best orbit shown previously in Figure~\ref{fig:orbit} is reproduced here for a visual comparison to the data. The model captures the observed profiles in $\mu_\alpha$, $\mu_\delta$ and $\varpi$, although some slight discrepancies towards the ends of the stream are apparent, possibly due to a mismatch between the real potential and the model used here. The heliocentric radial velocity as a function of Galactic longitude is shown on the bottom panel (this information is missing in the Gaia DR2 catalog), but the positions of the two stars of our sample that were fortuitously measured by the SDSS are shown in blue. The orbital path in Galactic coordinates of this best-fit model was shown previously in Figure~\ref{fig:sky_plots} (top panel).}
\label{fig:orbit-data}
\end{figure}

The fitting parameters are sky position $\alpha$, $\delta$, distance $d$, heliocentric velocity $v_h$ and proper motions $\mu_\alpha$, $\mu_\delta$. We chose to anchor the solutions at $\delta=-27\deg$, approximately half-way along the stream, leaving all other parameters free to be varied by the algorithm (obviously, without an anchor line, the solution would have wandered over the full length of the stream).

After rejecting a burn-in phase of $10^5$ steps, the MCMC procedure was run for a further $10^6$ iterations. 
The best-fit orbit is shown in Figure~\ref{fig:orbit}, integrated over a period of $2\Gyr$ in the same \citet{Dehnen:1998tk} model employed above for stream detection. The red region marks the part of the orbit where we have currently detected the structure. The orbit is disk-like, but strongly retrograde, possessing an apocenter at $R=19.8\pm0.3\kpc$, a pericenter at $4.94\pm0.01\kpc$ and a maximum height from the Galactic plane of $7.1\pm0.1\kpc$. As we show in Figure~\ref{fig:orbit-data}, the orbit fits the Gaia proper motions and parallax data very well, and we predict a very large heliocentric velocity gradient along the length of the stream (bottom panel). The physical length of the currently-observed portion of the stream is $\sim 5.3\kpc$.

\begin{figure}
\begin{center}
\includegraphics[angle=0, viewport= 10 10 405 390, clip, width=\hsize]{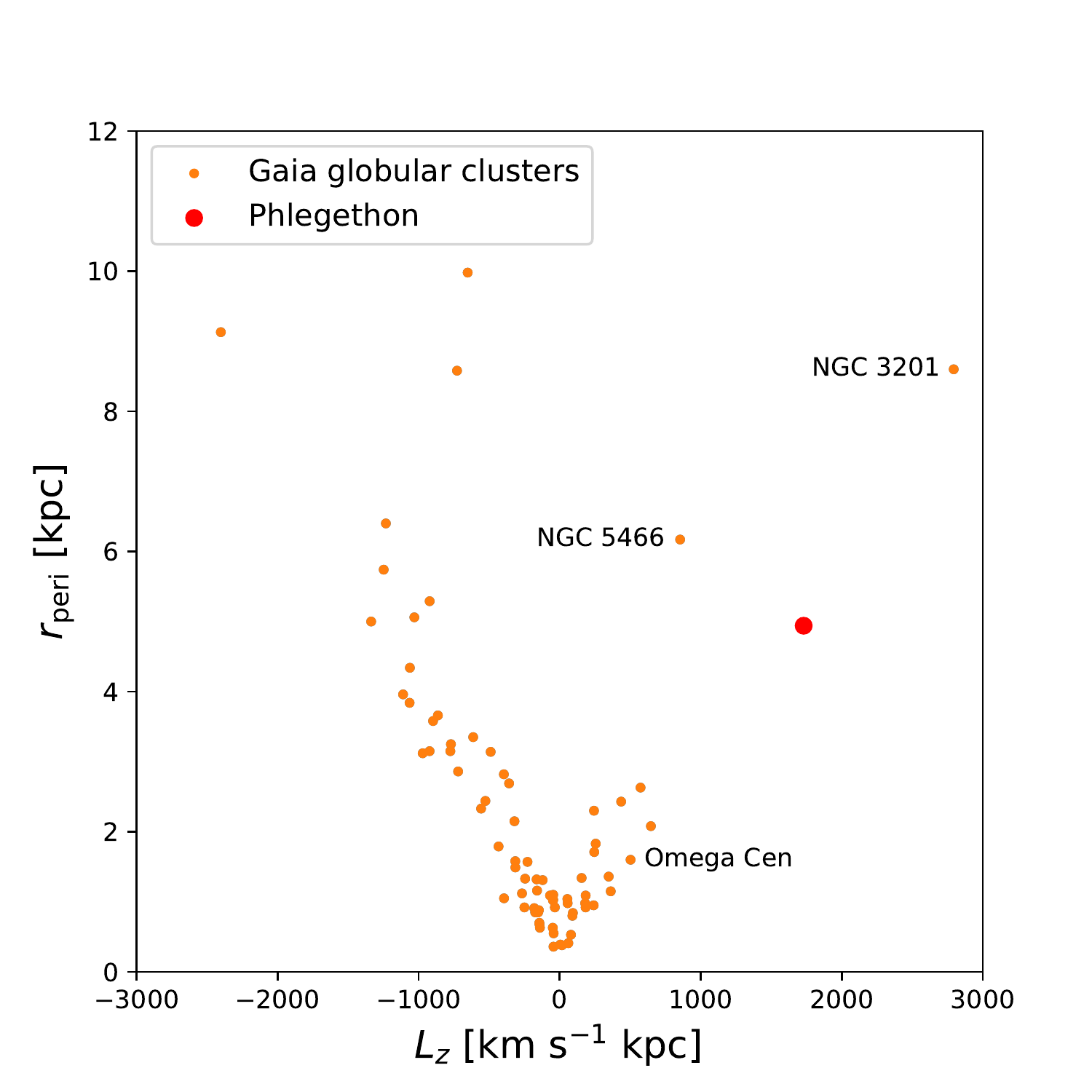}
\end{center}
\caption{Orbital properties of the Phlegethon stream compared to Milky Way globular clusters. The Galactic globular clusters with orbits measured by Gaia \citep{2018arXiv180409381G} are shown here in the $L_z$ $r_{\rm peri}$ plane. Retrograde objects have positive $L_z$. The properties of Phlegethon are shown by a large red dot: it can be seen to lie far from any other globular cluster, so we conclude that the progenitor system is now completely dissolved. The uncertainties on $L_z$ and $r_{\rm peri}$ derived here using the adopted \citet{Dehnen:1998tk} Galactic mass model are smaller than the size of the dot; however, the real uncertainties on these values are likely to be dominated by the reliability of our current knowledge of the mass distribution in the Milky Way, which Phlegethon may soon help to improve.}
\label{fig:GCs}
\end{figure}

\section{Discussion and Conclusions}
\label{sec:Discussion}

In this paper, we present the discovery of a new stream structure, that we named Phlegethon, found in the recently released Gaia DR2 dataset. Phlegethon, which lies close to the Sun at a mean heliocentric distance of $3.8\kpc$, possesses highly retrograde orbital motion. Two of the stars in our Gaia sample have measured radial velocities in the SDSS/Segue DR10 catalog; these objects confirm the reality of the new stream, and the additional information was used to improve slightly the orbital fits. The rms dispersion of the stream perpendicular to the fitted orbit is $1.4\deg$, corresponding to $88\pc$ at that heliocentric distance (for comparison, the Palomar~5 stream has a dispersion of $58\pc$, \citealt{2016ApJ...819....1I}). This implies that the Phlegethon stream must be the remnant of a disrupted globular cluster rather than a dwarf galaxy. 

To ascertain whether Phlegethon could be related to any surviving globular cluster, in Figure~\ref{fig:GCs} we compare the orbital properties ($z$-component of angular momentum $L_z$ and pericenter distance) of this stream with the globular cluster sample recently analyzed by \citet{2018arXiv180409381G}. We have labeled certain well-known retrograde globular clusters. Phlegethon clearly has extreme properties ($L_z=1769\pm8 \kms \kpc$ in the adopted model of the Galactic potential), with only NGC~3201 having a larger value of $L_z$. Most importantly, however, this analysis indicates that there are no known globular clusters with similar dynamical properties to this stream, so the progenitor has probably not survived (unless by chance it is hidden in the Galactic disk).

The very low mass of $2580\pm140\msun$ that we measure here, and the old stellar population age, raise an interesting puzzle. Clearly, dynamical friction must have been completely unimportant in affecting the orbit of the globular cluster progenitor, unless it arrived as part of a much more massive system, such as a dwarf galaxy. But the absence of a large population of stars on similar retrograde orbits suggests that this was not the case. Thus it would seem that the progenitor must have formed early in the life of the Milky Way, and continued orbiting in this disk-like but retrograde manner. At some point over the last $10\Gyr$ (the age of the stellar population as suggested by the CMD properties of the stream) the progenitor became unbound, and began forming the stream we have detected. It will be interesting to simulate this system to determine the dynamical age of the stream, and given its smooth appearance on the sky (Figure~\ref{fig:sky_plots}), examine the expected effect of heating by dark matter sub-halos and the repeated crossings every $\sim 200\, {\rm Myr}$ of the dense inner regions of the Galactic disk.

It may prove very rewarding to hunt over the sky for stars with similar orbital properties. This would increase the lever-arm for orbital-fit analyses, allowing one to better test Galactic mass models. It will also allow improved constraints to be placed on the initial mass of Phlegethon's progenitor. 

The detection of Phlegethon shows that the \texttt{STREAMFINDER} algorithm can be successfully employed to find even nearby structures of exceedingly low surface brightness ($\Sigma_G \sim 34.3 \, {\rm mag \, arcsec^{-2}}$). So far, in paper~II and in the present contribution we have analyzed only a small part of the parameter space of possible stream structures; our next efforts will focus on expanding the search criteria, and trying to explore further these fossil remnants in the Milky Way, examining their implications for galaxy formation and the dark matter problem.

\acknowledgments

This work has made use of data from the European Space Agency (ESA) mission {\it Gaia} (\url{https://www.cosmos.esa.int/gaia}), processed by the {\it Gaia} Data Processing and Analysis Consortium (DPAC, \url{https://www.cosmos.esa.int/web/gaia/dpac/consortium}). Funding for the DPAC has been provided by national institutions, in particular the institutions participating in the {\it Gaia} Multilateral Agreement. 

RAI and NFM gratefully acknowledge support from a ``Programme National Cosmologie et Galaxies'' grant. ES gratefully acknowledges funding by the Emmy Noether program from the Deutsche Forschungsgemeinschaft (DFG). RAI would like to thank his son Oliver for his help in naming this structure by pointing out that one of the five rivers of the Underworld was missing on previous stream maps \citep{2009ApJ...693.1118G}.

\bibliography{ms}

\begin{thebibliography}{31}
\expandafter\ifx\csname natexlab\endcsname\relax\def\natexlab#1{#1}\fi

\bibitem[{Bonaca \& Hogg(2018)}]{2018arXiv180406854B}
Bonaca, A., \& Hogg, D.~W. 2018, arXiv, arXiv:1804.06854

\bibitem[{Bovy {et~al.}(2016)Bovy, Bahmanyar, Fritz, \&
  Kallivayalil}]{2016ApJ...833...31B}
Bovy, J., Bahmanyar, A., Fritz, T.~K., \& Kallivayalil, N. 2016, ApJ, 833, 31

\bibitem[{Bressan {et~al.}(2012)Bressan, Marigo, Girardi, Salasnich, Dal~Cero,
  Rubele, \& Nanni}]{2012MNRAS.427..127B}
Bressan, A., Marigo, P., Girardi, L., Salasnich, B., Dal~Cero, C., Rubele, S.,
  \& Nanni, A. 2012, MNRAS, 427, 127

\bibitem[{Carlberg(2012)}]{2012ApJ...748...20C}
Carlberg, R.~G. 2012, ApJ, 748, 20

\bibitem[{Dehnen \& Binney(1998)}]{Dehnen:1998tk}
Dehnen, W., \& Binney, J. 1998, Royal Astronomical Society, 294, 429

\bibitem[{Dehnen {et~al.}(2004)Dehnen, Odenkirchen, Grebel, \&
  Rix}]{Dehnen:2004ez}
Dehnen, W., Odenkirchen, M., Grebel, E.~K., \& Rix, H.-W. 2004, AJ, 127, 2753

\bibitem[{Erkal {et~al.}(2016)Erkal, Belokurov, Bovy, \&
  Sanders}]{2016MNRAS.463..102E}
Erkal, D., Belokurov, V., Bovy, J., \& Sanders, J.~L. 2016, MNRAS, 463, 102

\bibitem[{{Gaia Collaboration} {et~al.}(2018{\natexlab{a}}){Gaia
  Collaboration}, Brown, Vallenari, Prusti, de~Bruijne, Babusiaux, \&
  Bailer-Jones}]{2018arXiv180409365G}
{Gaia Collaboration}, Brown, A. G.~A., Vallenari, A., Prusti, T., de~Bruijne,
  J. H.~J., Babusiaux, C., \& Bailer-Jones, C. A.~L. 2018{\natexlab{a}}, arXiv,
  arXiv:1804.09365

\bibitem[{{Gaia Collaboration} {et~al.}(2018{\natexlab{b}}){Gaia
  Collaboration}, Helmi, van Leeuwen, McMillan, Massari, Antoja, Robin,
  Lindegren, Bastian, \& {et al.}}]{2018arXiv180409381G}
{Gaia Collaboration} {et~al.} 2018{\natexlab{b}}, arXiv, arXiv:1804.09381

\bibitem[{{Gaia Collaboration} {et~al.}(2018{\natexlab{c}}){Gaia
  Collaboration}, Katz, Sartoretti, Cropper, Panuzzo, Seabroke, Viala, Benson,
  Blomme, \& {et al.}}]{2018arXiv180409372K}
---. 2018{\natexlab{c}}, arXiv, arXiv:1804.09372

\bibitem[{{Gaia Collaboration} {et~al.}(2018{\natexlab{d}}){Gaia
  Collaboration}, Lindegren, Hern{\'a}ndez, Bombrun, Klioner, Bastian,
  Ramos-Lerate, de~Torres, Steidelm{\"u}ller, \& {et
  al.}}]{2018arXiv180409366L}
---. 2018{\natexlab{d}}, arXiv, arXiv:1804.09366

\bibitem[{{Gaia Collaboration} {et~al.}(2016){Gaia Collaboration}, Prusti,
  de~Bruijne, Brown, Vallenari, Babusiaux, Bailer-Jones, Bastian, \& {et
  al.}}]{2016A&A...595A...1G}
---. 2016, A\&A, 595, A1

\bibitem[{Goodman \& Weare(2010)}]{Goodman:2010we}
Goodman, J., \& Weare, J. 2010, Commun. Appl. Math. Comput. Sci., 5, 65

\bibitem[{Grillmair(2009)}]{2009ApJ...693.1118G}
Grillmair, C.~J. 2009, ApJ, 693, 1118

\bibitem[{Grillmair \& Dionatos(2006)}]{2006ApJ...643L..17G}
Grillmair, C.~J., \& Dionatos, O. 2006, ApJ, 643, L17

\bibitem[{Ibata {et~al.}(2001)Ibata, Lewis, Irwin, Totten, \&
  Quinn}]{Ibata:2001be}
Ibata, R., Lewis, G.~F., Irwin, M., Totten, E., \& Quinn, T. 2001, ApJ, 551,
  294

\bibitem[{Ibata {et~al.}(2013)Ibata, Nipoti, Sollima, Bellazzini, Chapman, \&
  Dalessandro}]{2013MNRAS.428.3648I}
Ibata, R., Nipoti, C., Sollima, A., Bellazzini, M., Chapman, S.~C., \&
  Dalessandro, E. 2013, MNRAS, 428, 3648

\bibitem[{Ibata {et~al.}(2002)Ibata, Lewis, Irwin, \&
  Quinn}]{2002MNRAS.332..915I}
Ibata, R.~A., Lewis, G.~F., Irwin, M.~J., \& Quinn, T. 2002, MNRAS, 332, 915

\bibitem[{Ibata {et~al.}(2016)Ibata, Lewis, \& Martin}]{2016ApJ...819....1I}
Ibata, R.~A., Lewis, G.~F., \& Martin, N.~F. 2016, ApJ, 819, 1

\bibitem[{Johnston {et~al.}(2008)Johnston, Bullock, Sharma, Font, Robertson, \&
  Leitner}]{Johnston:2008jp}
Johnston, K.~V., Bullock, J.~S., Sharma, S., Font, A., Robertson, B.~E., \&
  Leitner, S.~N. 2008, ApJ, 689, 936

\bibitem[{Johnston {et~al.}(2002)Johnston, Spergel, \&
  Haydn}]{2002ApJ...570..656J}
Johnston, K.~V., Spergel, D.~N., \& Haydn, C. 2002, ApJ, 570, 656

\bibitem[{Karim \& Mamajek(2017)}]{2017MNRAS.465..472K}
Karim, M.~T., \& Mamajek, E.~E. 2017, MNRAS, 465, 472

\bibitem[{K{\"u}pper {et~al.}(2015)K{\"u}pper, Balbinot, Bonaca, Johnston,
  Hogg, Kroupa, \& Santiago}]{2015ApJ...803...80K}
K{\"u}pper, A. H.~W., Balbinot, E., Bonaca, A., Johnston, K.~V., Hogg, D.~W.,
  Kroupa, P., \& Santiago, B.~X. 2015, ApJ, 803, 80

\bibitem[{Law \& Majewski(2010)}]{2010ApJ...714..229L}
Law, D.~R., \& Majewski, S.~R. 2010, ApJ, 714, 229

\bibitem[{Malhan \& Ibata(2018)}]{2018MNRAS.477.4063M}
Malhan, K., \& Ibata, R. 2018, arXiv, 4063

\bibitem[{Malhan {et~al.}(2018)Malhan, Ibata, \& Martin}]{2018arXiv180411339M}
Malhan, K., Ibata, R.~A., \& Martin, N.~F. 2018, arXiv, arXiv:1804.11339

\bibitem[{Reid {et~al.}(2014)Reid, Menten, Brunthaler, Zheng, Dame, Xu, Wu,
  Zhang, \& {et al.}}]{2014ApJ...783..130R}
Reid, M.~J., {et~al.} 2014, ApJ, 783, 130

\bibitem[{Schlegel {et~al.}(1998)Schlegel, Finkbeiner, \&
  Davis}]{Schlegel:1998fw}
Schlegel, D.~J., Finkbeiner, D.~P., \& Davis, M. 1998, ApJ, 500, 525

\bibitem[{Sch{\"o}nrich {et~al.}(2010)Sch{\"o}nrich, Binney, \&
  Dehnen}]{2010MNRAS.403.1829S}
Sch{\"o}nrich, R., Binney, J., \& Dehnen, W. 2010, MNRAS, 403, 1829

\bibitem[{Varghese {et~al.}(2011)Varghese, Ibata, \&
  Lewis}]{2011MNRAS.417..198V}
Varghese, A., Ibata, R., \& Lewis, G.~F. 2011, MNRAS, 417, 198

\bibitem[{Yanny {et~al.}(2009)Yanny, Rockosi, Newberg, Knapp, Adelman-McCarthy,
  Alcorn, Allam, Allende~Prieto, \& {et al.}}]{2009AJ....137.4377Y}
Yanny, B., {et~al.} 2009, AJ, 137, 4377

\end{thebibliography}
\bibliographystyle{apj}

\end{document}